\documentclass[aps,prl,reprint,twocolumn,superscriptaddress,preprintnumbers,nofootinbib]{revtex4-1}

\usepackage{amsthm}
\usepackage{amsmath}
\usepackage{graphicx}
\usepackage{subcaption}
\usepackage{slashed}
\usepackage{amssymb}
\usepackage{float}
\usepackage[utf8]{inputenc}
\usepackage[T1]{fontenc}
\usepackage[hyperindex,breaklinks,colorlinks=True, citecolor=blue,urlcolor=blue,linkcolor=blue]{hyperref}
\usepackage[cal=cm,scr=euler]{mathalpha}
\usepackage[dvipsnames]{xcolor}
\usepackage[normalem]{ulem}
\usepackage{cancel}

\begin{document}

\title{Universal Equilibration Condition for Heavy Quarks}

\author{Krishna Rajagopal}
\email{krishna@mit.edu}
\affiliation{Center for Theoretical Physics -- a Leinweber Institute, Massachusetts Institute of Technology, Cambridge, MA 02139, USA}

\author{Bruno Scheihing-Hitschfeld}
\email{bscheihi@kitp.ucsb.edu}
\affiliation{Kavli Institute for Theoretical Physics, University of California, Santa Barbara, California 93106, USA}

\author{Urs Achim Wiedemann}
\email{urs.wiedemann@cern.ch}
\affiliation{Theoretical Physics Department, CERN, CH-1211 Gen\`eve 23, Switzerland}

\preprint{CERN-TH-2025-088, MIT-CTP/5858}

\date{\today}

\begin{abstract}
Kinetic equilibration at late times is physically required for heavy particles in a finite temperature medium. In Fokker-Planck dynamics, it is ensured by the Einstein relation between the drag and longitudinal momentum diffusion coefficients. However, in certain gauge field theories, this relation is violated at any nonzero heavy quark velocity. Recent work in strongly coupled $\mathcal{N}=4$ SYM gauge theory shows that the Kolmogorov equation for the heavy quark phase space distribution (that reduces to Fokker-Planck form upon truncating the momentum transfer probability distribution to second moments) does equilibrate even though the Fokker-Planck equation does not. 
Going beyond these (to date theory-specific) insights, we derive a
universal equilibration condition for the kernel of the Kolmogorov equation and, consequently, for the momentum transfer probability distribution that holds in any quantum field theory with any coupling strength.
This condition, which is the generalization of the Einstein relation to quantum field theories which feature non-Gaussian fluctuations,
reveals that the asymmetry between energy loss and energy gain in the momentum transfer probability distribution takes a simple, theory-independent, form.
\end{abstract}

\maketitle

In the kinetic theory of Brownian motion, the Einstein relation~\cite{Sutherland1905,Einstein1905,vSmoluchowski1906} is a linear relation 
\begin{equation}
	2 T p\, \eta_D = v\, \kappa_L\,  ,
    \label{eq:eq1}
\end{equation}
between the drag coefficient $\eta_D$ and the longitudinal momentum diffusion coefficient $\kappa_L$
for a particle of mass $M \gg T$, velocity $v$ and momentum $p$ embedded in a medium with temperature $T$. 
In the limit $v\to 0$, this relation is a direct consequence of the Fluctuation-Dissipation theorem~\cite{Callen1951,Kubo1966}. At nonzero velocity $v > 0$, Eq.~\eqref{eq:eq1} is not enforced by a general theorem, but it appears in the Langevin equation as the necessary and sufficient condition for the distribution of heavy particles to reach thermal equilibrium at late times~\cite{Langevin1908}. As such, it is employed as a consistency condition in Fokker-Planck/Langevin descriptions of heavy quarks in quark gluon plasma~\cite{Das:2013kea,Berrehrah:2014kba,Prino:2016cni,Xu:2017obm,Rapp:2018qla,Ke:2018tsh,Xu:2018gux,Cao:2018ews,Dong:2019unq,He:2022ywp}.  
However, for heavy quarks at finite velocity, the Einstein relation is violated. In  perturbation theory, Moore and Teaney~\cite{Moore:2004tg} demonstrated that at weak coupling the Einstein relation holds only up to leading log accuracy when $v > 0$.  
And, soon afterwards, when the drag~\cite{Herzog:2006gh,Gubser:2006bz} and momentum diffusion~\cite{Casalderrey-Solana:2006fio,Gubser:2006nz,Casalderrey-Solana:2007ahi} coefficients were first calculated in the strongly coupled limit of $\mathcal{N}=4$ Supersymmetric Yang-Mills (SYM) theory, Gubser~\cite{Gubser:2006nz} made the striking observation that the Einstein relation is badly broken, with $\tfrac{v \kappa_L}{2T p \eta_D} = \gamma^{3/2}$. This observation was further explored in Refs.~\cite{Giecold:2009cg,Casalderrey-Solana:2009ifi,Bu:2021jlp}, but remained puzzling. Exploiting more recent developments in the toolkit of the AdS/CFT 
correspondence~\cite{Skenderis:2008dg,Skenderis:2008dh,vanRees:2009rw,DEramo:2010wup}, in Ref.~\cite{Rajagopal:2025ukd} we demonstrated how this violation can be reconciled with the requirement that kinetic theory must drive heavy quarks to equilibrium in $\mathcal{N}=4$ SYM. Rather than characterizing the momentum transfer probability distribution $P({\bf k};{\bf v})$ of heavy quarks moving with velocity ${\bf v}$ in the ${\mathcal N}=4$ SYM plasma with temperature $T$ via its first and second moments $\eta_D$ and $\kappa_L$ only, working in the t’Hooft strong coupling limit and taking $M\gg T$ we obtained an analytic expression for the complete $P({\bf k};{\bf v})$ probability distribution, including expressions for all of its higher longitudinal, transverse and mixed moments. This enabled us to derive the Kolmogorov equation that evolves heavy quark distributions with knowledge of the full $P({\bf k};{\bf v})$. This equation has the Boltzmann distribution as a stationary solution, which is to say it does thermalize, even though truncating the Kolmogorov equation up to second moments in $P({\bf k};{\bf v})$ results in a Fokker-Planck equation that violates Einstein’s relation precisely as first found by Gubser~\cite{Gubser:2006nz}. 

In the strong coupling limit of $\mathcal{N}=4$ SYM, we thus have a complete understanding of how an interplay between all the higher moments of $P({\bf k};{\bf v})$ ensures kinetic equilibration of heavy quarks when Einstein’s relation is violated at finite velocity. These results raise questions about the general model-independent conditions for heavy quark thermalization in any quantum field theory: If a Fokker-Planck formulation of heavy quark equilibration fails at finite velocity because Einstein’s relation is violated, but a Kolmogorov formulation works, is there a general condition that the kernel of the Kolmogorov equation and, consequently, $P({\bf k};{\bf v})$ must satisfy? And, is this equilibration condition always satisfied at nonzero $v$, or could it be violated like Einstein’s relation in the Fokker-Planck formulation? Here, we provide answers to these questions.

Let us begin with our conclusions.
We shall indeed derive a universal equilibration condition, a generalization of the Einstein relation that is
always satisfied for a heavy quark ($M\gg T$) moving with any velocity ${\bf v}$ through the medium in thermal equilibrium at temperature $T$ described by any quantum field theory, whether weakly or strongly coupled, whether conformal or asymptotically free. We shall derive this universal condition first as a condition on the kernel of the Kolmogorov equation and then as a condition on the momentum transfer probability distribution $P({\bf k};{\bf v})$.
In the latter form, it implies a symmetry that 
is valid in any quantum field theory and independent of any model assumptions:
\begin{equation}
    P({\bf k};{\bf v}) \exp\left[- \frac{{\bf v}\cdot {\bf k}}{2T} \right] =
    P(-{\bf k};{\bf v}) \exp\left[ \frac{{\bf v}\cdot {\bf k}}{2T} \right]\, .
    \label{eq:eq32}
\end{equation}
In short, the asymmetry between momentum loss and momentum gain in the momentum transfer distribution $P({\bf k};{\bf v})$ is {\it always} given by the factor $\exp\left[ \frac{{\bf v}\cdot {\bf k}}{2T} \right]$. This nontrivial condition is in fact equivalent to having detailed balance in the corresponding stochastic process. We look forward to seeing Eq.~\eqref{eq:eq32} tested against data, as it provides a non-trivial consistency condition for any empirical determination of $P({\bf k};{\bf v})$.

Our starting point is the Heavy Quark Effective Theory (HQET) Lagrangian~\cite{Georgi:1990um}
\begin{equation}
    \mathcal{L}_{\rm HQET} = \bar{Q}_v i v^\mu D_\mu Q_{v} + \mathcal{O}(1/M) \, , \label{eq:L-HQET}
\end{equation}
where $v^\mu$ is the 4-velocity of the heavy quark, $Q_v$ is the heavy quark field describing small momentum fluctuations around its ``hard'' momentum $p^\mu = M v^\mu$, and $D_\mu = \partial_\mu - i g A_\mu$ is the covariant derivative. This form holds for any spin of the heavy particle and any gauge group. At leading (zeroth) order in $1/M$, the heavy quark propagator is the Wilson line 
\begin{equation}
    W^{ij}_{[x_f,x_i]} = P \exp \left( i g \int_{t_i}^{t_f} dt \,  A_{\mu} \dot{x}^\mu \right) \, . \label{eq:wilson-line}
\end{equation}
Along the path $x^\mu(t) = x_i^\mu + (t-t_i, {\bf v} (t-t_i))$ up to $x_f = x^\mu(t_f)$, the color of the heavy quark rotates in the position-dependent gauge field $A_\mu$ from $i$ to $j$, and the heavy quark momentum changes by an amount ${\bf k}$ according to the probability amplitude
\begin{equation}
    \langle {\bf p} - {\bf k},i |_{\rm out} | {\bf p},j \rangle_{\rm in} = \int d^3{\bf x}_f e^{i {\bf k} \cdot {\bf x}_f } W^{ij}_{[x_f,x_i]} \, . \label{eq:eq4}
\end{equation}
Here ${\bf p} = M \gamma {\bf v} $, and our sign convention for ${\bf k}$ is that ${\bf k} \cdot {\bf p} > 0$ corresponds to loss of longitudinal momentum. Following Born's rule, taking the absolute value squared of this amplitude and averaging it over the thermal density matrix $e^{-\beta H}$ with $H$ the Hamiltonian of the gauge theory that describes the medium and with $\beta = 1/T$, one finds the probability distribution for the momentum ${\bf k}$  transferred from the heavy quark to the medium as it travels through it with velocity ${\bf v}$ for a time $t$: 
\begin{align}
    P({\bf k};{\bf v}) &= \frac{1}{(2\pi)^3} \int d^3{\bf L} \, e^{-i {\bf k} \cdot {\bf L}  } \langle W_{\bf v} \rangle ({\bf L}) \, , \label{eq:P-from-W}
\end{align}
where the Wilson lines enter through
\begin{align}
    &\langle W_{\bf v} \rangle ({\bf L}) \label{eq:Wloop-Wlines} \\
    &=\frac{1}{Z} {\rm Tr}_{\mathcal{H}} \! \left[ W^{lk}_{[(0, {\bf L}),(t, {\bf v} t + {\bf L} )]} \tilde{\rho}_{ki} W^{ij}_{[(t, {\bf v} t ),(0,{\bf 0})]} e^{-\beta H}  \rho_{jl} \right] \, . 
    \nonumber
\end{align}
As the color indices attached to the Wilson lines end at different spatial positions in the amplitude and complex conjugate amplitude, they need to be parallel transported 
by colored operators $\rho_{jl}$ at the initial time and $\tilde{\rho}_{ki}$ at the final time to yield a gauge-invariant expression. These operators $\rho$, $\tilde{\rho}$ encode information about the heavy quark initial state preparation and final state measurement. Having in mind that it is possible (but not necessary)
for $\rho$, $\tilde{\rho}$ to be Wilson lines, we refer to $\langle W_{\bf v} \rangle ({\bf L})$ as a Wilson loop. The matrices $\tilde{\rho}$, $\rho$ depend on ${\bf L}$, but not on $t$. The normalization factor $Z$ is chosen such that $\langle W_{\bf v}\rangle ({\bf 0}) = 1$ --- it is independent of ${\bf L}$ but may depend on $t$ and ${\bf v}$.

If $t \gg 1/T, |{\bf L}|$, it is convenient to write the Wilson loop as
\begin{equation}
    \langle W_{\bf v} \rangle ({\bf L}) = \exp \left( - t T S({\bf L}; {\bf v}) + \ldots \right) \, , \label{eq:W-S}
\end{equation}
which serves to define $S$ and where ``$\ldots$'' represents terms that are not extensive with $t$, i.e., they are of order $\mathcal{O}( (tT)^{-n} )$ with $n \geq 0$. Higher powers of $t$ cannot appear in the exponent of this asymptotic expansion, since this would imply that $\tfrac{\partial_t \langle W_{\bf v} \rangle}{\langle W_{\bf v} \rangle} $ depends on the total amount of time passed, which would violate time translation invariance. Explicit calculations at strong coupling~\cite{Rajagopal:2025ukd} and general diagrammatic arguments~\cite{DEramo:2012uzl} confirm this.  

One is typically interested in the probability distribution $\mathscr{P}({\bf p}, \tau)$ of the {\it total} heavy quark momentum. The evolution of this distribution in a finite temperature plasma can be updated with knowledge of the momentum {\it transfer} probability $P({\bf k};{\bf v})$. After a time $\Delta t$, small enough so that the probability for the momentum to have been significantly modified is negligible, we have
\begin{align}
    &\mathscr{P}({\bf p}, \tau + \Delta t) \\
    &= \int d^3 {\bf p}' P(  {\bf p}' - {\bf p} ; {\bf v} = {\bf v}({\bf p}') , t = \Delta t ) \mathscr{P}({\bf p}',\tau) \, . \nonumber
\end{align}
The Kolmogorov evolution equation for $\mathscr{P}({\bf p}, \tau)$ is obtained by
differentiating this equation with respect to $\Delta t$, using Eqs.~\eqref{eq:P-from-W} and~\eqref{eq:W-S} and integrating by parts, yielding
\begin{align}
    \partial_\tau \mathscr{P} = - T K(\partial_{\bf p}, {\bf p}) \mathscr{P} \, ,
    \label{eq:Kolmogorov}
\end{align}
where we have defined
\begin{align}
    K({\bf x}, {\bf p}) \equiv S\! \left( -i {\bf x} ; \, {\bf v} = {\bf v}({\bf p}) \right) \, , \label{eq:K-operator}
\end{align}
the kernel of the Kolmogorov equation. Eq.~\eqref{eq:Kolmogorov} takes the most general form possible for a continuous-time Markov process. Deviations from this form, encoded in the subleading non-extensive $\mathcal{O}((tT)^{-n})$ terms in Eq.~\eqref{eq:W-S}, characterize short-duration non-Markovian ``memory effects'' in the dynamics. 
If $\Delta t \gg 1/T$, non-extensive terms are negligible and the dynamics is a Markov process~\cite{Breuer:2007juk}.
Expanding Eq.~\eqref{eq:Kolmogorov} in powers of $\partial_{\bf p}$, as Kramers and Moyal first did in the 1940s~\cite{Kramers1940,Moyal1949}, and truncating this expansion to second order yields the Fokker-Planck equation.

To fully specify Eq.~\eqref{eq:Kolmogorov}, it is necessary to provide the dispersion relation of the heavy particle $E = E({\bf p})$, which determines its velocity of propagation ${\bf v}({\bf p}) = \partial E/\partial{\bf p}$. Note that we have introduced ${\bf x} = i{\bf L}$ in order to make it clear that no imaginary numbers appear in Eq.~\eqref{eq:Kolmogorov}. In Eq.~\eqref{eq:Kolmogorov}, the derivatives in the first argument of $K$ act on its second argument as well as $\mathscr{P}$. That is to say, in a power series representation, these derivatives are inserted to the left of the rest of the expression in each term. 

We now come to the main aims of the present work which is to demonstrate that for any quantum field theory, the Kolmogorov kernel \eqref{eq:K-operator} satisfies properties that ensure heavy quark equilibration at late times and that constrain the form of the momentum transfer probability $P({\bf k};{\bf v})$. We shall do so by exploiting that $K$ inherits its structure from the Wilson loop $\langle W_{\bf v} \rangle({\bf L})$ which is well-defined in any quantum field theory.

Upon deriving the explicit result for the full evolution kernel $K({\bf x}, {\bf p})$ in ${\mathcal N}=4$ SYM theory in Ref.~\cite{Rajagopal:2025ukd}, we had noted there that, even though truncating the kernel up to second moments leads to a Fokker-Planck equation that does not equilibrate, the 
full Kolmogorov Eq.~\eqref{eq:Kolmogorov} has as a stationary solution the equilibrium distribution 
\begin{equation}
    \mathcal{P}({\bf p}) \propto \exp \left( - \frac{E({\bf p})}{T} \right) \, .
    \label{eq:eq14}
\end{equation}
To generalize these insights to any quantum field theory, we first note that this equilibrium condition should always hold. The right-hand side of Eq.~\eqref{eq:Kolmogorov} must vanish on physical grounds if acting on \eqref{eq:eq14}. As we work only to leading order in $T/M$, this equilibrium condition simplifies since one needs to consider only a subset of all possible terms generated by acting on $\exp \left( - \frac{E({\bf p})}{T} \right)$ with the operator $K(\partial_{\bf p}, {\bf p})$. To see this, note that $\tfrac{\partial E}{\partial {\bf p}} = {\bf v}$, and therefore after a derivative acts on $E({\bf p})$, the next one can only act on another $E({\bf p})$ or on ${\bf v}$. Also, for any massive particle, the derivatives of ${\bf v}$ with respect to ${\bf p}$ are suppressed by powers of $1/M$. This means that if we only keep track of the leading (zeroth) order terms in $T/M$, the evolution equation Eq.~\eqref{eq:Kolmogorov} may be equivalently rewritten as
\begin{equation}
    \partial_\tau \ln \mathscr{P} = - T K \big( \partial_{\bf p} \ln \mathscr{P}, {\bf p} \big) \, , \label{eq:Kolmogorov-leading}
\end{equation}
where the normalization prefactor of the distribution is to be calculated from $\ln \mathscr{P}$ at each point in time to conserve heavy quark number, which is conserved because of the structure of Eq.~\eqref{eq:Kolmogorov}. This is consistent because the dynamics of this normalization prefactor appears at subleading order in $T/M$ in Eq.~\eqref{eq:Kolmogorov}.  It follows from Eq.~\eqref{eq:Kolmogorov-leading} that at leading order in $T/M$ the Kolmogorov equation equilibrates to \eqref{eq:eq14} if and only if
\begin{equation}
    S( i {\bf v}/T ; {\bf v} ) = 0 \, . \label{eq:equil-cond}
\end{equation}
This is the equilibration condition for the kernel of the Kolmogorov equation,
$K(- {\bf v}/T , {\bf p}) = 0$. Note that it is independent of the precise form of $E({\bf p})$.

Before deriving the condition on the momentum transfer probability distribution $P({\bf k};{\bf v})$ implied by \eqref{eq:equil-cond}, we shall show that the equilibration condition \eqref{eq:equil-cond} is a universal property that is always satisfied for any value of $v$ on general field-theoretic grounds. Starting from Eq.~\eqref{eq:Wloop-Wlines} and dropping the normalization $Z$ for notational convenience, we find
\begin{align}
    &\langle W_{\bf v} \rangle ({\bf L}) \nonumber \\
    &= 
    {\rm Tr}\! \left[ \tilde{\rho}_{ik}^{\mathcal{P}\mathcal{T} }  W^{kl}_{[(-t, -{\bf v} t - {\bf L} ),(0, -{\bf L})]}  \rho_{lj}^{\mathcal{P}\mathcal{T} }  e^{-\beta H} W^{ji}_{[(0,{\bf 0}),(-t, -{\bf v} t )]}  \right] \nonumber \\
    &=
    {\rm Tr}\! \left[ \tilde{\rho}_{ik}^{\mathcal{P}\mathcal{T} }  W^{kl}_{[(-t, -{\bf v} t - {\bf L} ),(0, -{\bf L})]}  \rho_{lj}^{\mathcal{P}\mathcal{T} } W^{ji}_{[(i\beta,{\bf 0}),(i\beta-t, -{\bf v} t )]} e^{-\beta H}  \right] \nonumber \\
    &= 
    {\rm Tr}\! \left[   W^{kl}_{[(-i\beta, - {\bf L} ),(t - i\beta, {\bf v} t - {\bf L})]}  \rho_{lj}^{\mathcal{P}\mathcal{T} } W^{ji}_{[(t,{\bf v}t ),(0, {\bf 0} )]} e^{-\beta H} \tilde{\rho}_{ik}^{\mathcal{P}\mathcal{T} } \right] \nonumber \\
    &= 
    {\rm Tr}\! \left[ W^{kl}_{[(0, - {\bf L} + i\beta {\bf v} ),(t , {\bf v} t - {\bf L} + i\beta {\bf v} )]}  \rho_{lj}^{\mathcal{P}\mathcal{T} } W^{ji}_{[(t,{\bf v}t ),(0, {\bf 0} )]} e^{-\beta H} \tilde{\rho}_{ik}^{\mathcal{P}\mathcal{T} } \right] \nonumber \\
    &= \langle W_{\bf v} \rangle (-{\bf L} + i {\bf v}/T ) \, . \label{eq:W-KMS}
\end{align}
Here, the second line follows from applying parity $\mathcal{P}$ and time reversal $\mathcal{T}$ transformations to \eqref{eq:Wloop-Wlines}. The matrices $\rho, \tilde{\rho}$ get transformed to their $\mathcal{P}\mathcal{T}$ conjugates in this step. We then insert $e^{\beta H} e^{-\beta H}$ to the left of the closing bracket (3rd line) and we translate in the 4th line by $(i\beta + t, {\bf v}t)$ (this also affects $\rho, \tilde{\rho}$, but we omit this in the notation). The 5th line relies on $t\gg 1/T$. It follows from 
\begin{equation}
    W^{kl}_{[(t_0, {\bf L} ),(t + t_0, {\bf v} t + {\bf L} ) ]} = e^{i Px} W^{kl}_{[(0, {\bf 0} ),(t , {\bf v} t  ) ]} e^{-i Px} \, ,
\end{equation}
where $P^\mu$ is the 4-momentum operator (the generator of spacetime translations) and $x^\mu = (t_0, {\bf L})$ is an analytic function of $t_0$ and each of the individual components $L_i$. In the limit of an infinitely long line (justified by $t \gg 1/T$), $x^\mu = (t_0, {\bf L})$ possesses a certain reparametrization invariance for real $t_0$, ${\bf L}$: the operator only depends on the combination ${\bf L} - {\bf v} t_0$. Because the function is analytic, it follows that this property gets extended into the complex plane, and so the imaginary time shift by $-i\beta$ can be moved to the position argument. We emphasize once more that the relations~\eqref{eq:W-KMS} are equalities only for the $t$-extensive contribution, on which we focus throughout. 

It follows immediately from Eq.~\eqref{eq:W-KMS} that
\begin{equation}
    S({\bf L};{\bf v}) = S(-{\bf L} + i {\bf v}/T;{\bf v}) \, , \label{eq:S-analytic-property}
\end{equation}
independently of the details of the theory. Combined with the normalization $S({\bf 0};{\bf v}) = 0$, it follows that $S(i {\bf v}/T;{\bf v}) = 0$. This is our first main result: the kinetic equilibrium condition \eqref{eq:equil-cond} is a universal property that the Wilson loop \eqref{eq:W-KMS} satisfies in any quantum field theory, resulting from an analog of a KMS condition~\cite{Kubo:1957mj,Martin:1959jp}. 
The result~\eqref{eq:S-analytic-property} has immediate consequences beyond guaranteeing the existence of kinetic equilibrium as a solution to the Kolmogorov equation, including consequences for the form that the momentum transfer distribution $P({\bf k};{\bf v})$ must take.
The symmetry property \eqref{eq:eq32} of 
$P({\bf k};{\bf v})$ with which we began
follows directly from Eq.~\eqref{eq:W-KMS} by Fourier transforming $\langle W_{\bf v} \rangle ({\bf L}) = \langle W_{\bf v} \rangle (-{\bf L} + i{\bf v}/T )$, which yields $P({\bf k};{\bf v}) = e^{ {\bf v} \cdot {\bf k}/T } P(-{\bf k};{\bf v})$.

We now present a second argument that leads
to the same conclusion about the symmetry property of $P({\bf k}; {\bf v})$  and that makes important aspects of our conclusions more explicit. 
In the Markovian limit $\Delta t \gg 1/T$, we may carry out the integral in Eq.~\eqref{eq:P-from-W} by means of the saddle point approximation. Omitting terms that do not scale exponentially with $t$, one finds that
\begin{equation}
    P({\bf k};{\bf v}) = \exp \left( - t T \tilde{S}( {\bf C} ; {\bf v} ) + \ldots \right) \, , \label{eq:P-H}
\end{equation}
where ${\bf C}\equiv {\bf k}/(tT)$ and where $\tilde S$ is a Legendre transform of $S$. 
The normalization of the distribution \eqref{eq:P-H} is enforced by the subleading terms. Since
a heavy quark propagating through the plasma will lose momentum  as a linear function of time, it is natural to use in the following the intensive momentum transfer rate in units of the temperature, ${\bf C}$, rather than ${\bf k}$.
The logarithm of the Wilson loop $S({\bf L};{\bf v})$ is related to the logarithm of the transfer probability $\tilde{S}({\bf C};{\bf v})$ by a Legendre transform, $\tilde{S}({\bf C};{\bf v}) = \left[ S({\bf L};{\bf v}) - {\bf L} \cdot \frac{\partial S}{\partial {\bf L} } \right]_{{\bf L} = {\bf L}({\bf C}) }$ with   ${\bf C} = i \frac{\partial S}{\partial {\bf L}}$~\cite{Rajagopal:2025ukd}. 
Therefore
\begin{equation}
    \tilde{S}({\bf C};{\bf v}) = \left[ K({\bf x},{\bf p}) - {\bf x} \cdot \frac{\partial K}{\partial {\bf x}} \right]_{{\bf x} = {\bf x}({\bf C})} \, , \label{eq:Legendre-tf}
\end{equation}
where ${\bf x}({\bf C})$ is determined by
\begin{equation}
    {\bf C} = - \frac{\partial K}{\partial {\bf x} } \, . \label{eq:x-of-C}
\end{equation}
If this equation admits more than one solution, the dominant saddle is to be taken.
Eq.~\eqref{eq:S-analytic-property} implies
\begin{align}
    K({\bf x},{\bf p}) = K(-{\bf x} - {\bf v}/T, {\bf p}) \, , \label{eq:K-symm}
\end{align}
which means that $K$ is symmetric about ${\bf x} = -{\bf v}/(2T)$. To make this symmetry property apparent, we introduce
\begin{align}
    {\bf x}' &\equiv {\bf x} + \frac{\bf v}{2T} \, , \\
    \tilde{K}({\bf x}',{\bf p}) &\equiv K({\bf x}' - {\bf v}/(2T) ,{\bf p}) = K({\bf x},{\bf p}) \, ,
\end{align}
with which $\tilde{K}({\bf x}',{\bf p}) = \tilde{K}(-{\bf x}',{\bf p})$ and the momentum transfer rate is ${\bf C} = - \partial \tilde{K} / \partial{\bf x}' $. It follows that
\begin{eqnarray}
    \tilde{S}({\bf C};{\bf v}) &=& \left[ \tilde{K}({\bf x}',{\bf p}) - \left({\bf x}' -  \frac{\bf v}{2T} \right) \cdot \frac{\partial \tilde{K}}{\partial {\bf x}'} \right]_{{\bf x}' = {\bf x}'({\bf C})} \nonumber \\
    &=& -\frac{{\bf v} \cdot {\bf C} }{2T} + 
    \underbrace{\left[ \tilde{K}({\bf x}',{\bf p}) - {\bf x}' \cdot \frac{\partial \tilde{K}}{\partial {\bf x}'} \right]_{{\bf x}' = {\bf x}'({\bf C})} }_{= \tilde{S}_e({\bf C};{\bf v}) } \!\! \! \! \! \! \! \!\! \! \!   , \label{eq:H-result}
\end{eqnarray}
making it apparent that $\tilde{S}$ is determined by $K$.

The symmetry property~\eqref{eq:eq32} of the momentum transfer distribution $P({\bf k};{\bf v})$ also follows directly from Eq.~\eqref{eq:H-result}.
Since $\tilde{K}$ is an even function of ${\bf x}'$, the even part $\tilde{S}_{e}({\bf C};{\bf v}) = \tilde{S}_{e}(-{\bf C};{\bf v})$ of $\tilde{S}({\bf C};{\bf v})$ is theory-dependent while the odd part is always exactly equal to $-{\bf v} \cdot {\bf C}/(2T)$, in any quantum field theory. 
(Note that the drag force that a heavy quark experiences is determined by the minimum of $\tilde{S}$, i.e., the value of ${\bf C}$ that satisfies $\tfrac{\partial \tilde{S}_{e}}{\partial{\bf C}} = \tfrac{ {\bf v} }{2T}$, which implies $\tilde{S}_{e}({\bf C};{\bf v}) = \tfrac{ {\bf v} \cdot {\bf C} }{2T}$ if $S({\bf 0};{\bf v}) = 0$.) 
We have thus derived that the probability distribution for the momentum transferred from a heavy particle to a finite temperature plasma as it propagates through this medium is inherently asymmetric, 
\begin{equation}
    P({\bf k};{\bf v}) = \exp \left( - t  T \tilde{S}_{e}({\bf C};{\bf v}) + t \, {\bf v} \cdot {\bf C}/2 \right) \, ,
    \label{eq:prob}
\end{equation}
the explicit connection between $\tilde{S}_e$ and $K$, and that 
the asymmetry between energy loss (${\bf C}>0$) and energy gain (${\bf C}<0$) is model-independent.  
Eq.~\eqref{eq:prob} is the condition that $P({\bf k};{\bf v})$  must satisfy in order to guarantee the existence of kinetic equilibration as a solution to the Kolmogorov equation, and is thus the generalization of the Einstein relation. If the probability distribution $P({\bf k};{\bf v})$ were Gaussian with $\tilde S_e({\bf C};{\bf v})= \frac{T}{2 \kappa_L(v)} C_L^2 + \frac{T}{2\kappa_T(v)} C_\perp^2$, 
then (noting that the drag coefficient $\eta_D$ is defined via $\langle k_L \rangle = t \,p \,\eta_D$) the condition \eqref{eq:prob} would  imply the relation \eqref{eq:eq1} between its first and second moments. In any quantum field theory, Eq.~\eqref{eq:prob} is a condition on {\it all} the moments of $P({\bf k};{\bf v})$.

In the strongly coupled $\mathcal{N}=4$ SYM theory plasma, we have derived the explicit form for symmetric and asymmetric part of $\tilde S({\bf C};{\bf v})$~\cite{Rajagopal:2025ukd}. The asymmetric term $-{\bf v} \cdot {\bf C}/(2T)$ in \eqref{eq:prob} was derived as a contribution coming from the point where the worldsheet dual to the heavy quark touches the horizon of the spacetime dual to $\mathcal{N}=4$ SYM at finite $T$. Here, we understand it to arise as a model-independent universal feature.

Remarkably, this feature is also a sufficient condition for equilibration: if one is given a momentum transfer probability of the form of Eq.~\eqref{eq:prob}, then the Legendre transform of $\tilde{S}({\bf C};{\bf v}) = \tilde{S}_e({\bf C};{\bf v}) - \tfrac{{\bf v} \cdot {\bf C}}{2T}$, which by definition is equal to $K({\bf x},{\bf p})$ with ${\bf x} = \tfrac{\partial \tilde{S}_e}{\partial {\bf C} } - \tfrac{\bf v}{2T} $, is automatically an even function of ${\bf x} + \tfrac{\bf v}{2T}$, and therefore Eq.~\eqref{eq:K-symm} and Eq.~\eqref{eq:S-analytic-property} follow. 

The KMS-like condition \eqref{eq:S-analytic-property} is thus equivalent to the statement that the momentum transfer distribution $P({\bf k};{\bf v})$ takes the 
form \eqref{eq:prob} with the symmetry \eqref{eq:eq32}. It is pleasing to see that the detailed balance equation for a Markov process with a stationary Boltzmann distribution
\begin{align}
    &\exp \left(-\frac{E({\bf p})}{T} \right) P({\bf k};{\bf v}({\bf p}) ) \nonumber \\ 
    &= \exp \left(-\frac{E({\bf p} - {\bf k})}{T} \right) P(-{\bf k};{\bf v}({\bf p} - {\bf k}) ) \, ,
\end{align}
is automatically satisfied at the first non-trivial order in ${\bf k}/M$. 
The consequences of the form of $P({\bf k};{\bf v})$ can also be expressed in terms of the connected moments of $P({\bf k};{\bf v})$. To explore these, we start from Eq.~\eqref{eq:equil-cond} and note that $S({\bf L};{\bf v})$ is the generating functional of the  connected moments of $P({\bf k};{\bf v})$ which can be written as~\cite{Rajagopal:2025ukd} 
\begin{align}
    S({\bf L};{\bf v}) &= -  \sum_{m,n=0}^\infty
    \frac{i^{2m+n}}{(2m)!n!} 
    \frac{\langle k_T^{2m} k_L^n\rangle_c}{t\,T}
    \nonumber \\
    &\qquad \quad \times
    \left({\bf L}^2 - \left(\frac{{\bf v} \cdot {\bf L} }{v}  \right)^2 \right)^m \!\! \left(\frac{{\bf v} \cdot {\bf L}}{v} \right)^n\!\! ,
    \label{eq:fullS}
\end{align}
where $k_T$ and $k_L$ are one-dimensional projections of ${\bf k}$ along ${\bf v}$ and orthogonal to it, with $v \equiv |{\bf v}|$.
For ${\bf L} = i {\bf v}/T$ as in Eq.~\eqref{eq:equil-cond}, the transverse projector in \eqref{eq:fullS} vanishes. Therefore, evaluating \eqref{eq:fullS} for the equilibrium condition $S(i {\bf v}/T;{\bf v}) = 0$, one finds
\begin{equation}
    \sum_{n=1}^\infty \frac{(-v)^n}{n!} 
    c_n(v) = 0 \, ,
    \label{eq:csum}
\end{equation}
where $c_n(v) \equiv \langle k_L^n \rangle_c / (t \, T^{n+1})$. In Eq.~\eqref{eq:csum}, the contribution for $n=0$ vanishes since $S({\bf L}=0;{\bf v})=0$. Various key consequences of \eqref{eq:equil-cond} are then immediate:
\begin{enumerate}
    \item The particle does not get dragged if it is not moving: 
\begin{equation}
    c_1(0)=0\, .
\end{equation}
    
    \item The fluctuation-dissipation theorem  
    \begin{equation}
        c_2(0) = 2 \lim_{v \to 0} c_1(v)/v \, 
                \label{eq:eq29}
    \end{equation}
    holds.
    \item If and only if higher order moments vanish ($c_n(v) = 0$ for $n > 2$), the Einstein relation 
    \begin{equation}
        2 c_1(v) = v\, c_2(v)\, 
                \label{eq:eq30}
    \end{equation}
    holds.
    \item In general, the higher moments will not vanish. To analyze the general case, it is helpful to expand $c_n(v)$ as a power series in $v$. If $c_n(v) = \sum_{k=0}^{\infty} d_{n,k} v^k$, then for every positive integer $m$
    \begin{equation}
        \sum_{l=1}^m \frac{(-1)^l}{l!} d_{l,m-l} = 0 \, .
        \label{eq:eq31}
    \end{equation}
\end{enumerate}
These results show from first-principles how the Einstein relation \eqref{eq:eq30} arises in the special case in which the momentum transfer probability distribution is Gaussian. In general, for any quantum field theory, the generalization Eq.~\eqref{eq:prob} of the Einstein equation is a condition on {\it all} the moments of $P({\bf k};{\bf v})$, and it is an interplay between all the moments, not just the first two, that guarantees equilibration. Correspondingly, one can use Eq.~\eqref{eq:eq31} to extract information about the higher-order moments of $P({\bf k},{\bf v})$ from the lower order ones. If one supplements the above with further knowledge about the $v$-dependence of the coefficients, more stringent conditions can be derived. 

We conclude by relating the present study to other recent developments in heavy quark transport. In response to the above-mentioned limitations of Fokker-Planck dynamics and to ensure kinetic equilibration for arbitrary velocity of the heavy quark, several groups have studied Boltzmann transport formulations and used them to benchmark earlier Langevin treatments~\cite{Das:2013kea,Xu:2017obm,Ke:2018tsh,Rapp:2018qla,Cao:2018ews}. 
As of today, these provide the most detailed phenomenological description of heavy quark parton energy loss and flow phenomena measured in ultra-relativistic heavy-ion collisions~\cite{Das:2013kea,Berrehrah:2014kba,Prino:2016cni,Xu:2017obm,Ke:2018tsh,Xu:2018gux,Rapp:2018qla,Dong:2019unq,Cao:2018ews,He:2022ywp}. 
However, the validity of Boltzmann transport is limited to weakly coupled plasmas modeled via quasiparticles, and the choice of the Boltzmann collision kernel typically involves 
further model-dependent assumptions which make it difficult to relate the Boltzmann collision kernel to  transport properties calculated from first principles in a thermal field theory. 

Like Boltzmann transport, the range of applicability of the Kolmogorov equation that we have introduced has no limitation regarding the heavy quark velocity, and its late-time evolution is guaranteed to equilibrate. 
Unlike Boltzmann transport, it makes no assumptions regarding whether the theory is weakly or strongly coupled, or whether the plasma is composed of quasiparticles. Moreover, we have shown that the Kolmogorov kernel $K$ is connected directly to quantities calculated from first principles in quantum field theory. Indeed, the symmetry property $K({\bf x}, {\bf p}) = K(-{\bf x}-{\bf v}/T, {\bf p})$ implies that its series expansion around ${\bf x} = -{\bf v}/T$ (equilibrium) has the same expansion coefficients (up to an alternating sign) as its series around ${\bf x} = 0$, which are in turn determined by the moments of $P({\bf k}; {\bf v})$. The connection between $K$ and moments of $P({\bf k}; {\bf v})$ is important since the low moments of $P({\bf k};{\bf v})$ such as the heavy quark diffusion coefficient related to $\kappa_L$ in the $v\rightarrow 0$ limit can be calculated directly in finite temperature field theory at weak coupling~\cite{Svetitsky:1987gq,Moore:2004tg,Caron-Huot:2007rwy,Caron-Huot:2008dyw} and strong coupling~\cite{Casalderrey-Solana:2006fio,Gubser:2006nz,Casalderrey-Solana:2007ahi} as well as via lattice techniques~\cite{Petreczky:2005nh,Caron-Huot:2009ncn,Banerjee:2011ra,Brambilla:2020siz,Altenkort:2020fgs,Brambilla:2022xbd,Altenkort:2023oms,Altenkort:2023eav}. 
While a full analytic calculation of the evolution operator $K$ is not yet feasible in QCD 
(though it has been achieved in strongly coupled $\mathcal{N}=4$ SYM~\cite{Rajagopal:2025ukd}), it may nevertheless be constrained by knowledge of the low-order moments of the momentum transfer distribution $P({\bf k}; {\bf v})$, obtained either via {\it ab initio} quantum field theory calculations or via comparison with experiment. Constraining these heavy quark transport properties is a central aim of future heavy-ion
collision experiments during the high-luminosity phase of the CERN LHC  program~\cite{Adamova:2019vkf}. We expect that the Kolmogorov equation~\eqref{eq:Kolmogorov}, due to its particularly clear relation with these transport coefficients, will be an asset for establishing this link between current and  future data and fundamental properties of finite temperature QCD.

\begin{acknowledgments}
We are grateful to Xi Dong, Guy Moore, and Robinson Mancilla for very helpful discussions. This work was supported in part by the U.S.~Department of Energy, Office of Science, Office of Nuclear Physics under grant Contract Number DE-SC0011090\@, by grant NSF PHY-2309135 to the Kavli Institute for Theoretical Physics (KITP), and by grant 994312 from the Simons Foundation.
\end{acknowledgments}

\bibliography{main-arXiv-v2.bib}

\begin{thebibliography}{49}%
\makeatletter
\providecommand \@ifxundefined [1]{%
 \@ifx{#1\undefined}
}%
\providecommand \@ifnum [1]{%
 \ifnum #1\expandafter \@firstoftwo
 \else \expandafter \@secondoftwo
 \fi
}%
\providecommand \@ifx [1]{%
 \ifx #1\expandafter \@firstoftwo
 \else \expandafter \@secondoftwo
 \fi
}%
\providecommand \natexlab [1]{#1}%
\providecommand \enquote  [1]{``#1''}%
\providecommand \bibnamefont  [1]{#1}%
\providecommand \bibfnamefont [1]{#1}%
\providecommand \citenamefont [1]{#1}%
\providecommand \href@noop [0]{\@secondoftwo}%
\providecommand \href [0]{\begingroup \@sanitize@url \@href}%
\providecommand \@href[1]{\@@startlink{#1}\@@href}%
\providecommand \@@href[1]{\endgroup#1\@@endlink}%
\providecommand \@sanitize@url [0]{\catcode `\\12\catcode `\$12\catcode `\&12\catcode `\#12\catcode `\^12\catcode `\_12\catcode `\%12\relax}%
\providecommand \@@startlink[1]{}%
\providecommand \@@endlink[0]{}%
\providecommand \url  [0]{\begingroup\@sanitize@url \@url }%
\providecommand \@url [1]{\endgroup\@href {#1}{\urlprefix }}%
\providecommand \urlprefix  [0]{URL }%
\providecommand \Eprint [0]{\href }%
\providecommand \doibase [0]{http://dx.doi.org/}%
\providecommand \selectlanguage [0]{\@gobble}%
\providecommand \bibinfo  [0]{\@secondoftwo}%
\providecommand \bibfield  [0]{\@secondoftwo}%
\providecommand \translation [1]{[#1]}%
\providecommand \BibitemOpen [0]{}%
\providecommand \bibitemStop [0]{}%
\providecommand \bibitemNoStop [0]{.\EOS\space}%
\providecommand \EOS [0]{\spacefactor3000\relax}%
\providecommand \BibitemShut  [1]{\csname bibitem#1\endcsname}%
\let\auto@bib@innerbib\@empty
\bibitem [{\citenamefont {Sutherland}(1905)}]{Sutherland1905}%
  \BibitemOpen
  \bibfield  {author} {\bibinfo {author} {\bibfnamefont {W.}~\bibnamefont {Sutherland}},\ }\href {\doibase 10.1080/14786440509463331} {\bibfield  {journal} {\bibinfo  {journal} {The London, Edinburgh, and Dublin Philosophical Magazine and Journal of Science}\ }\textbf {\bibinfo {volume} {9}},\ \bibinfo {pages} {781} (\bibinfo {year} {1905})}\BibitemShut {NoStop}%
\bibitem [{\citenamefont {Einstein}(1905)}]{Einstein1905}%
  \BibitemOpen
  \bibfield  {author} {\bibinfo {author} {\bibfnamefont {A.}~\bibnamefont {Einstein}},\ }\href {\doibase 10.1002/andp.19053220806} {\bibfield  {journal} {\bibinfo  {journal} {Annalen der Physik}\ }\textbf {\bibinfo {volume} {322}},\ \bibinfo {pages} {549} (\bibinfo {year} {1905})}\BibitemShut {NoStop}%
\bibitem [{\citenamefont {von Smoluchowski}(1906)}]{vSmoluchowski1906}%
  \BibitemOpen
  \bibfield  {author} {\bibinfo {author} {\bibfnamefont {M.}~\bibnamefont {von Smoluchowski}},\ }\href {\doibase 10.1002/andp.19063261405} {\bibfield  {journal} {\bibinfo  {journal} {Annalen der Physik}\ }\textbf {\bibinfo {volume} {326}},\ \bibinfo {pages} {756} (\bibinfo {year} {1906})}\BibitemShut {NoStop}%
\bibitem [{\citenamefont {Callen}\ and\ \citenamefont {Welton}(1951)}]{Callen1951}%
  \BibitemOpen
  \bibfield  {author} {\bibinfo {author} {\bibfnamefont {H.~B.}\ \bibnamefont {Callen}}\ and\ \bibinfo {author} {\bibfnamefont {T.~A.}\ \bibnamefont {Welton}},\ }\href {\doibase 10.1103/PhysRev.83.34} {\bibfield  {journal} {\bibinfo  {journal} {Phys. Rev.}\ }\textbf {\bibinfo {volume} {83}},\ \bibinfo {pages} {34} (\bibinfo {year} {1951})}\BibitemShut {NoStop}%
\bibitem [{\citenamefont {Kubo}(1966)}]{Kubo1966}%
  \BibitemOpen
  \bibfield  {author} {\bibinfo {author} {\bibfnamefont {R.}~\bibnamefont {Kubo}},\ }\href {\doibase 10.1088/0034-4885/29/1/306} {\bibfield  {journal} {\bibinfo  {journal} {Reports on Progress in Physics}\ }\textbf {\bibinfo {volume} {29}},\ \bibinfo {pages} {255} (\bibinfo {year} {1966})}\BibitemShut {NoStop}%
\bibitem [{\citenamefont {Langevin}(1908)}]{Langevin1908}%
  \BibitemOpen
  \bibfield  {author} {\bibinfo {author} {\bibfnamefont {P.}~\bibnamefont {Langevin}},\ }\href@noop {} {\bibfield  {journal} {\bibinfo  {journal} {C. R. Acad. Sci. (Paris), 1908, 146, 530}\ } (\bibinfo {year} {1908})}\BibitemShut {NoStop}%
\bibitem [{\citenamefont {Das}\ \emph {et~al.}(2014)\citenamefont {Das}, \citenamefont {Scardina}, \citenamefont {Plumari},\ and\ \citenamefont {Greco}}]{Das:2013kea}%
  \BibitemOpen
  \bibfield  {author} {\bibinfo {author} {\bibfnamefont {S.~K.}\ \bibnamefont {Das}}, \bibinfo {author} {\bibfnamefont {F.}~\bibnamefont {Scardina}}, \bibinfo {author} {\bibfnamefont {S.}~\bibnamefont {Plumari}}, \ and\ \bibinfo {author} {\bibfnamefont {V.}~\bibnamefont {Greco}},\ }\href {\doibase 10.1103/PhysRevC.90.044901} {\bibfield  {journal} {\bibinfo  {journal} {Phys. Rev. C}\ }\textbf {\bibinfo {volume} {90}},\ \bibinfo {pages} {044901} (\bibinfo {year} {2014})},\ \Eprint {http://arxiv.org/abs/1312.6857} {arXiv:1312.6857 [nucl-th]} \BibitemShut {NoStop}%
\bibitem [{\citenamefont {Berrehrah}\ \emph {et~al.}(2014)\citenamefont {Berrehrah}, \citenamefont {Gossiaux}, \citenamefont {Aichelin}, \citenamefont {Cassing},\ and\ \citenamefont {Bratkovskaya}}]{Berrehrah:2014kba}%
  \BibitemOpen
  \bibfield  {author} {\bibinfo {author} {\bibfnamefont {H.}~\bibnamefont {Berrehrah}}, \bibinfo {author} {\bibfnamefont {P.-B.}\ \bibnamefont {Gossiaux}}, \bibinfo {author} {\bibfnamefont {J.}~\bibnamefont {Aichelin}}, \bibinfo {author} {\bibfnamefont {W.}~\bibnamefont {Cassing}}, \ and\ \bibinfo {author} {\bibfnamefont {E.}~\bibnamefont {Bratkovskaya}},\ }\href {\doibase 10.1103/PhysRevC.90.064906} {\bibfield  {journal} {\bibinfo  {journal} {Phys. Rev. C}\ }\textbf {\bibinfo {volume} {90}},\ \bibinfo {pages} {064906} (\bibinfo {year} {2014})},\ \Eprint {http://arxiv.org/abs/1405.3243} {arXiv:1405.3243 [hep-ph]} \BibitemShut {NoStop}%
\bibitem [{\citenamefont {Prino}\ and\ \citenamefont {Rapp}(2016)}]{Prino:2016cni}%
  \BibitemOpen
  \bibfield  {author} {\bibinfo {author} {\bibfnamefont {F.}~\bibnamefont {Prino}}\ and\ \bibinfo {author} {\bibfnamefont {R.}~\bibnamefont {Rapp}},\ }\href {\doibase 10.1088/0954-3899/43/9/093002} {\bibfield  {journal} {\bibinfo  {journal} {J. Phys. G}\ }\textbf {\bibinfo {volume} {43}},\ \bibinfo {pages} {093002} (\bibinfo {year} {2016})},\ \Eprint {http://arxiv.org/abs/1603.00529} {arXiv:1603.00529 [nucl-ex]} \BibitemShut {NoStop}%
\bibitem [{\citenamefont {Xu}\ \emph {et~al.}(2018)\citenamefont {Xu}, \citenamefont {Bernhard}, \citenamefont {Bass}, \citenamefont {Nahrgang},\ and\ \citenamefont {Cao}}]{Xu:2017obm}%
  \BibitemOpen
  \bibfield  {author} {\bibinfo {author} {\bibfnamefont {Y.}~\bibnamefont {Xu}}, \bibinfo {author} {\bibfnamefont {J.~E.}\ \bibnamefont {Bernhard}}, \bibinfo {author} {\bibfnamefont {S.~A.}\ \bibnamefont {Bass}}, \bibinfo {author} {\bibfnamefont {M.}~\bibnamefont {Nahrgang}}, \ and\ \bibinfo {author} {\bibfnamefont {S.}~\bibnamefont {Cao}},\ }\href {\doibase 10.1103/PhysRevC.97.014907} {\bibfield  {journal} {\bibinfo  {journal} {Phys. Rev. C}\ }\textbf {\bibinfo {volume} {97}},\ \bibinfo {pages} {014907} (\bibinfo {year} {2018})},\ \Eprint {http://arxiv.org/abs/1710.00807} {arXiv:1710.00807 [nucl-th]} \BibitemShut {NoStop}%
\bibitem [{\citenamefont {Beraudo}\ \emph {et~al.}(2018)\citenamefont {Beraudo} \emph {et~al.}}]{Rapp:2018qla}%
  \BibitemOpen
  \bibfield  {author} {\bibinfo {author} {\bibfnamefont {A.}~\bibnamefont {Beraudo}} \emph {et~al.},\ }\href {\doibase 10.1016/j.nuclphysa.2018.09.002} {\bibfield  {journal} {\bibinfo  {journal} {Nucl. Phys. A}\ }\textbf {\bibinfo {volume} {979}},\ \bibinfo {pages} {21} (\bibinfo {year} {2018})},\ \Eprint {http://arxiv.org/abs/1803.03824} {arXiv:1803.03824 [nucl-th]} \BibitemShut {NoStop}%
\bibitem [{\citenamefont {Ke}\ \emph {et~al.}(2018)\citenamefont {Ke}, \citenamefont {Xu},\ and\ \citenamefont {Bass}}]{Ke:2018tsh}%
  \BibitemOpen
  \bibfield  {author} {\bibinfo {author} {\bibfnamefont {W.}~\bibnamefont {Ke}}, \bibinfo {author} {\bibfnamefont {Y.}~\bibnamefont {Xu}}, \ and\ \bibinfo {author} {\bibfnamefont {S.~A.}\ \bibnamefont {Bass}},\ }\href {\doibase 10.1103/PhysRevC.98.064901} {\bibfield  {journal} {\bibinfo  {journal} {Phys. Rev. C}\ }\textbf {\bibinfo {volume} {98}},\ \bibinfo {pages} {064901} (\bibinfo {year} {2018})},\ \Eprint {http://arxiv.org/abs/1806.08848} {arXiv:1806.08848 [nucl-th]} \BibitemShut {NoStop}%
\bibitem [{\citenamefont {Xu}\ \emph {et~al.}(2019)\citenamefont {Xu} \emph {et~al.}}]{Xu:2018gux}%
  \BibitemOpen
  \bibfield  {author} {\bibinfo {author} {\bibfnamefont {Y.}~\bibnamefont {Xu}} \emph {et~al.},\ }\href {\doibase 10.1103/PhysRevC.99.014902} {\bibfield  {journal} {\bibinfo  {journal} {Phys. Rev. C}\ }\textbf {\bibinfo {volume} {99}},\ \bibinfo {pages} {014902} (\bibinfo {year} {2019})},\ \Eprint {http://arxiv.org/abs/1809.10734} {arXiv:1809.10734 [nucl-th]} \BibitemShut {NoStop}%
\bibitem [{\citenamefont {Cao}\ \emph {et~al.}(2019)\citenamefont {Cao} \emph {et~al.}}]{Cao:2018ews}%
  \BibitemOpen
  \bibfield  {author} {\bibinfo {author} {\bibfnamefont {S.}~\bibnamefont {Cao}} \emph {et~al.},\ }\href {\doibase 10.1103/PhysRevC.99.054907} {\bibfield  {journal} {\bibinfo  {journal} {Phys. Rev. C}\ }\textbf {\bibinfo {volume} {99}},\ \bibinfo {pages} {054907} (\bibinfo {year} {2019})},\ \Eprint {http://arxiv.org/abs/1809.07894} {arXiv:1809.07894 [nucl-th]} \BibitemShut {NoStop}%
\bibitem [{\citenamefont {Dong}\ and\ \citenamefont {Greco}(2019)}]{Dong:2019unq}%
  \BibitemOpen
  \bibfield  {author} {\bibinfo {author} {\bibfnamefont {X.}~\bibnamefont {Dong}}\ and\ \bibinfo {author} {\bibfnamefont {V.}~\bibnamefont {Greco}},\ }\href {\doibase 10.1016/j.ppnp.2018.08.001} {\bibfield  {journal} {\bibinfo  {journal} {Prog. Part. Nucl. Phys.}\ }\textbf {\bibinfo {volume} {104}},\ \bibinfo {pages} {97} (\bibinfo {year} {2019})}\BibitemShut {NoStop}%
\bibitem [{\citenamefont {He}\ \emph {et~al.}(2023)\citenamefont {He}, \citenamefont {van Hees},\ and\ \citenamefont {Rapp}}]{He:2022ywp}%
  \BibitemOpen
  \bibfield  {author} {\bibinfo {author} {\bibfnamefont {M.}~\bibnamefont {He}}, \bibinfo {author} {\bibfnamefont {H.}~\bibnamefont {van Hees}}, \ and\ \bibinfo {author} {\bibfnamefont {R.}~\bibnamefont {Rapp}},\ }\href {\doibase 10.1016/j.ppnp.2023.104020} {\bibfield  {journal} {\bibinfo  {journal} {Prog. Part. Nucl. Phys.}\ }\textbf {\bibinfo {volume} {130}},\ \bibinfo {pages} {104020} (\bibinfo {year} {2023})},\ \Eprint {http://arxiv.org/abs/2204.09299} {arXiv:2204.09299 [hep-ph]} \BibitemShut {NoStop}%
\bibitem [{\citenamefont {Moore}\ and\ \citenamefont {Teaney}(2005)}]{Moore:2004tg}%
  \BibitemOpen
  \bibfield  {author} {\bibinfo {author} {\bibfnamefont {G.~D.}\ \bibnamefont {Moore}}\ and\ \bibinfo {author} {\bibfnamefont {D.}~\bibnamefont {Teaney}},\ }\href {\doibase 10.1103/PhysRevC.71.064904} {\bibfield  {journal} {\bibinfo  {journal} {Phys. Rev. C}\ }\textbf {\bibinfo {volume} {71}},\ \bibinfo {pages} {064904} (\bibinfo {year} {2005})},\ \Eprint {http://arxiv.org/abs/hep-ph/0412346} {arXiv:hep-ph/0412346} \BibitemShut {NoStop}%
\bibitem [{\citenamefont {Herzog}\ \emph {et~al.}(2006)\citenamefont {Herzog}, \citenamefont {Karch}, \citenamefont {Kovtun}, \citenamefont {Kozcaz},\ and\ \citenamefont {Yaffe}}]{Herzog:2006gh}%
  \BibitemOpen
  \bibfield  {author} {\bibinfo {author} {\bibfnamefont {C.~P.}\ \bibnamefont {Herzog}}, \bibinfo {author} {\bibfnamefont {A.}~\bibnamefont {Karch}}, \bibinfo {author} {\bibfnamefont {P.}~\bibnamefont {Kovtun}}, \bibinfo {author} {\bibfnamefont {C.}~\bibnamefont {Kozcaz}}, \ and\ \bibinfo {author} {\bibfnamefont {L.~G.}\ \bibnamefont {Yaffe}},\ }\href {\doibase 10.1088/1126-6708/2006/07/013} {\bibfield  {journal} {\bibinfo  {journal} {JHEP}\ }\textbf {\bibinfo {volume} {07}},\ \bibinfo {pages} {013} (\bibinfo {year} {2006})},\ \Eprint {http://arxiv.org/abs/hep-th/0605158} {arXiv:hep-th/0605158} \BibitemShut {NoStop}%
\bibitem [{\citenamefont {Gubser}(2006)}]{Gubser:2006bz}%
  \BibitemOpen
  \bibfield  {author} {\bibinfo {author} {\bibfnamefont {S.~S.}\ \bibnamefont {Gubser}},\ }\href {\doibase 10.1103/PhysRevD.74.126005} {\bibfield  {journal} {\bibinfo  {journal} {Phys. Rev. D}\ }\textbf {\bibinfo {volume} {74}},\ \bibinfo {pages} {126005} (\bibinfo {year} {2006})},\ \Eprint {http://arxiv.org/abs/hep-th/0605182} {arXiv:hep-th/0605182} \BibitemShut {NoStop}%
\bibitem [{\citenamefont {Casalderrey-Solana}\ and\ \citenamefont {Teaney}(2006)}]{Casalderrey-Solana:2006fio}%
  \BibitemOpen
  \bibfield  {author} {\bibinfo {author} {\bibfnamefont {J.}~\bibnamefont {Casalderrey-Solana}}\ and\ \bibinfo {author} {\bibfnamefont {D.}~\bibnamefont {Teaney}},\ }\href {\doibase 10.1103/PhysRevD.74.085012} {\bibfield  {journal} {\bibinfo  {journal} {Phys. Rev. D}\ }\textbf {\bibinfo {volume} {74}},\ \bibinfo {pages} {085012} (\bibinfo {year} {2006})},\ \Eprint {http://arxiv.org/abs/hep-ph/0605199} {arXiv:hep-ph/0605199} \BibitemShut {NoStop}%
\bibitem [{\citenamefont {Gubser}(2008)}]{Gubser:2006nz}%
  \BibitemOpen
  \bibfield  {author} {\bibinfo {author} {\bibfnamefont {S.~S.}\ \bibnamefont {Gubser}},\ }\href {\doibase 10.1016/j.nuclphysb.2007.09.017} {\bibfield  {journal} {\bibinfo  {journal} {Nucl. Phys. B}\ }\textbf {\bibinfo {volume} {790}},\ \bibinfo {pages} {175} (\bibinfo {year} {2008})},\ \Eprint {http://arxiv.org/abs/hep-th/0612143} {arXiv:hep-th/0612143} \BibitemShut {NoStop}%
\bibitem [{\citenamefont {Casalderrey-Solana}\ and\ \citenamefont {Teaney}(2007)}]{Casalderrey-Solana:2007ahi}%
  \BibitemOpen
  \bibfield  {author} {\bibinfo {author} {\bibfnamefont {J.}~\bibnamefont {Casalderrey-Solana}}\ and\ \bibinfo {author} {\bibfnamefont {D.}~\bibnamefont {Teaney}},\ }\href {\doibase 10.1088/1126-6708/2007/04/039} {\bibfield  {journal} {\bibinfo  {journal} {JHEP}\ }\textbf {\bibinfo {volume} {04}},\ \bibinfo {pages} {039} (\bibinfo {year} {2007})},\ \Eprint {http://arxiv.org/abs/hep-th/0701123} {arXiv:hep-th/0701123} \BibitemShut {NoStop}%
\bibitem [{\citenamefont {Giecold}\ \emph {et~al.}(2009)\citenamefont {Giecold}, \citenamefont {Iancu},\ and\ \citenamefont {Mueller}}]{Giecold:2009cg}%
  \BibitemOpen
  \bibfield  {author} {\bibinfo {author} {\bibfnamefont {G.~C.}\ \bibnamefont {Giecold}}, \bibinfo {author} {\bibfnamefont {E.}~\bibnamefont {Iancu}}, \ and\ \bibinfo {author} {\bibfnamefont {A.~H.}\ \bibnamefont {Mueller}},\ }\href {\doibase 10.1088/1126-6708/2009/07/033} {\bibfield  {journal} {\bibinfo  {journal} {JHEP}\ }\textbf {\bibinfo {volume} {07}},\ \bibinfo {pages} {033} (\bibinfo {year} {2009})},\ \Eprint {http://arxiv.org/abs/0903.1840} {arXiv:0903.1840 [hep-th]} \BibitemShut {NoStop}%
\bibitem [{\citenamefont {Casalderrey-Solana}\ \emph {et~al.}(2009)\citenamefont {Casalderrey-Solana}, \citenamefont {Kim},\ and\ \citenamefont {Teaney}}]{Casalderrey-Solana:2009ifi}%
  \BibitemOpen
  \bibfield  {author} {\bibinfo {author} {\bibfnamefont {J.}~\bibnamefont {Casalderrey-Solana}}, \bibinfo {author} {\bibfnamefont {K.-Y.}\ \bibnamefont {Kim}}, \ and\ \bibinfo {author} {\bibfnamefont {D.}~\bibnamefont {Teaney}},\ }\href {\doibase 10.1088/1126-6708/2009/12/066} {\bibfield  {journal} {\bibinfo  {journal} {JHEP}\ }\textbf {\bibinfo {volume} {12}},\ \bibinfo {pages} {066} (\bibinfo {year} {2009})},\ \Eprint {http://arxiv.org/abs/0908.1470} {arXiv:0908.1470 [hep-th]} \BibitemShut {NoStop}%
\bibitem [{\citenamefont {Bu}\ and\ \citenamefont {Zhang}(2021)}]{Bu:2021jlp}%
  \BibitemOpen
  \bibfield  {author} {\bibinfo {author} {\bibfnamefont {Y.}~\bibnamefont {Bu}}\ and\ \bibinfo {author} {\bibfnamefont {B.}~\bibnamefont {Zhang}},\ }\href {\doibase 10.1103/PhysRevD.104.086002} {\bibfield  {journal} {\bibinfo  {journal} {Phys. Rev. D}\ }\textbf {\bibinfo {volume} {104}},\ \bibinfo {pages} {086002} (\bibinfo {year} {2021})},\ \Eprint {http://arxiv.org/abs/2108.10060} {arXiv:2108.10060 [hep-th]} \BibitemShut {NoStop}%
\bibitem [{\citenamefont {Skenderis}\ and\ \citenamefont {van Rees}(2009)}]{Skenderis:2008dg}%
  \BibitemOpen
  \bibfield  {author} {\bibinfo {author} {\bibfnamefont {K.}~\bibnamefont {Skenderis}}\ and\ \bibinfo {author} {\bibfnamefont {B.~C.}\ \bibnamefont {van Rees}},\ }\href {\doibase 10.1088/1126-6708/2009/05/085} {\bibfield  {journal} {\bibinfo  {journal} {JHEP}\ }\textbf {\bibinfo {volume} {05}},\ \bibinfo {pages} {085} (\bibinfo {year} {2009})},\ \Eprint {http://arxiv.org/abs/0812.2909} {arXiv:0812.2909 [hep-th]} \BibitemShut {NoStop}%
\bibitem [{\citenamefont {Skenderis}\ and\ \citenamefont {van Rees}(2008)}]{Skenderis:2008dh}%
  \BibitemOpen
  \bibfield  {author} {\bibinfo {author} {\bibfnamefont {K.}~\bibnamefont {Skenderis}}\ and\ \bibinfo {author} {\bibfnamefont {B.~C.}\ \bibnamefont {van Rees}},\ }\href {\doibase 10.1103/PhysRevLett.101.081601} {\bibfield  {journal} {\bibinfo  {journal} {Phys. Rev. Lett.}\ }\textbf {\bibinfo {volume} {101}},\ \bibinfo {pages} {081601} (\bibinfo {year} {2008})},\ \Eprint {http://arxiv.org/abs/0805.0150} {arXiv:0805.0150 [hep-th]} \BibitemShut {NoStop}%
\bibitem [{\citenamefont {van Rees}(2009)}]{vanRees:2009rw}%
  \BibitemOpen
  \bibfield  {author} {\bibinfo {author} {\bibfnamefont {B.~C.}\ \bibnamefont {van Rees}},\ }\href {\doibase 10.1016/j.nuclphysbps.2009.07.078} {\bibfield  {journal} {\bibinfo  {journal} {Nucl. Phys. B Proc. Suppl.}\ }\textbf {\bibinfo {volume} {192-193}},\ \bibinfo {pages} {193} (\bibinfo {year} {2009})},\ \Eprint {http://arxiv.org/abs/0902.4010} {arXiv:0902.4010 [hep-th]} \BibitemShut {NoStop}%
\bibitem [{\citenamefont {D'Eramo}\ \emph {et~al.}(2011)\citenamefont {D'Eramo}, \citenamefont {Liu},\ and\ \citenamefont {Rajagopal}}]{DEramo:2010wup}%
  \BibitemOpen
  \bibfield  {author} {\bibinfo {author} {\bibfnamefont {F.}~\bibnamefont {D'Eramo}}, \bibinfo {author} {\bibfnamefont {H.}~\bibnamefont {Liu}}, \ and\ \bibinfo {author} {\bibfnamefont {K.}~\bibnamefont {Rajagopal}},\ }\href {\doibase 10.1103/PhysRevD.84.065015} {\bibfield  {journal} {\bibinfo  {journal} {Phys. Rev. D}\ }\textbf {\bibinfo {volume} {84}},\ \bibinfo {pages} {065015} (\bibinfo {year} {2011})},\ \Eprint {http://arxiv.org/abs/1006.1367} {arXiv:1006.1367 [hep-ph]} \BibitemShut {NoStop}%
\bibitem [{\citenamefont {Rajagopal}\ \emph {et~al.}(2025)\citenamefont {Rajagopal}, \citenamefont {Scheihing-Hitschfeld},\ and\ \citenamefont {Wiedemann}}]{Rajagopal:2025ukd}%
  \BibitemOpen
  \bibfield  {author} {\bibinfo {author} {\bibfnamefont {K.}~\bibnamefont {Rajagopal}}, \bibinfo {author} {\bibfnamefont {B.}~\bibnamefont {Scheihing-Hitschfeld}}, \ and\ \bibinfo {author} {\bibfnamefont {U.~A.}\ \bibnamefont {Wiedemann}},\ }\href@noop {} {\  (\bibinfo {year} {2025})},\ \Eprint {http://arxiv.org/abs/2501.06289} {arXiv:2501.06289 [hep-ph]} \BibitemShut {NoStop}%
\bibitem [{\citenamefont {Georgi}(1990)}]{Georgi:1990um}%
  \BibitemOpen
  \bibfield  {author} {\bibinfo {author} {\bibfnamefont {H.}~\bibnamefont {Georgi}},\ }\href {\doibase 10.1016/0370-2693(90)91128-X} {\bibfield  {journal} {\bibinfo  {journal} {Phys. Lett. B}\ }\textbf {\bibinfo {volume} {240}},\ \bibinfo {pages} {447} (\bibinfo {year} {1990})}\BibitemShut {NoStop}%
\bibitem [{\citenamefont {D'Eramo}\ \emph {et~al.}(2013)\citenamefont {D'Eramo}, \citenamefont {Lekaveckas}, \citenamefont {Liu},\ and\ \citenamefont {Rajagopal}}]{DEramo:2012uzl}%
  \BibitemOpen
  \bibfield  {author} {\bibinfo {author} {\bibfnamefont {F.}~\bibnamefont {D'Eramo}}, \bibinfo {author} {\bibfnamefont {M.}~\bibnamefont {Lekaveckas}}, \bibinfo {author} {\bibfnamefont {H.}~\bibnamefont {Liu}}, \ and\ \bibinfo {author} {\bibfnamefont {K.}~\bibnamefont {Rajagopal}},\ }\href {\doibase 10.1007/JHEP05(2013)031} {\bibfield  {journal} {\bibinfo  {journal} {JHEP}\ }\textbf {\bibinfo {volume} {05}},\ \bibinfo {pages} {031} (\bibinfo {year} {2013})},\ \Eprint {http://arxiv.org/abs/1211.1922} {arXiv:1211.1922 [hep-ph]} \BibitemShut {NoStop}%
\bibitem [{\citenamefont {Breuer}\ and\ \citenamefont {Petruccione}(2007)}]{Breuer:2007juk}%
  \BibitemOpen
  \bibfield  {author} {\bibinfo {author} {\bibfnamefont {H.-P.}\ \bibnamefont {Breuer}}\ and\ \bibinfo {author} {\bibfnamefont {F.}~\bibnamefont {Petruccione}},\ }\href {\doibase 10.1093/acprof:oso/9780199213900.001.0001} {\emph {\bibinfo {title} {{The Theory of Open Quantum Systems}}}}\ (\bibinfo  {publisher} {Oxford University Press},\ \bibinfo {year} {2007})\BibitemShut {NoStop}%
\bibitem [{\citenamefont {Kramers}(1940)}]{Kramers1940}%
  \BibitemOpen
  \bibfield  {author} {\bibinfo {author} {\bibfnamefont {H.}~\bibnamefont {Kramers}},\ }\href {\doibase https://doi.org/10.1016/S0031-8914(40)90098-2} {\bibfield  {journal} {\bibinfo  {journal} {Physica}\ }\textbf {\bibinfo {volume} {7}},\ \bibinfo {pages} {284} (\bibinfo {year} {1940})}\BibitemShut {NoStop}%
\bibitem [{\citenamefont {Moyal}(1949)}]{Moyal1949}%
  \BibitemOpen
  \bibfield  {author} {\bibinfo {author} {\bibfnamefont {J.~E.}\ \bibnamefont {Moyal}},\ }\href {\doibase 10.1111/j.2517-6161.1949.tb00030.x} {\bibfield  {journal} {\bibinfo  {journal} {Journal of the Royal Statistical Society: Series B (Methodological)}\ }\textbf {\bibinfo {volume} {11}},\ \bibinfo {pages} {150} (\bibinfo {year} {1949})}\BibitemShut {NoStop}%
\bibitem [{\citenamefont {Kubo}(1957)}]{Kubo:1957mj}%
  \BibitemOpen
  \bibfield  {author} {\bibinfo {author} {\bibfnamefont {R.}~\bibnamefont {Kubo}},\ }\href {\doibase 10.1143/JPSJ.12.570} {\bibfield  {journal} {\bibinfo  {journal} {J. Phys. Soc. Jap.}\ }\textbf {\bibinfo {volume} {12}},\ \bibinfo {pages} {570} (\bibinfo {year} {1957})}\BibitemShut {NoStop}%
\bibitem [{\citenamefont {Martin}\ and\ \citenamefont {Schwinger}(1959)}]{Martin:1959jp}%
  \BibitemOpen
  \bibfield  {author} {\bibinfo {author} {\bibfnamefont {P.~C.}\ \bibnamefont {Martin}}\ and\ \bibinfo {author} {\bibfnamefont {J.~S.}\ \bibnamefont {Schwinger}},\ }\href {\doibase 10.1103/PhysRev.115.1342} {\bibfield  {journal} {\bibinfo  {journal} {Phys. Rev.}\ }\textbf {\bibinfo {volume} {115}},\ \bibinfo {pages} {1342} (\bibinfo {year} {1959})}\BibitemShut {NoStop}%
\bibitem [{\citenamefont {Svetitsky}(1988)}]{Svetitsky:1987gq}%
  \BibitemOpen
  \bibfield  {author} {\bibinfo {author} {\bibfnamefont {B.}~\bibnamefont {Svetitsky}},\ }\href {\doibase 10.1103/PhysRevD.37.2484} {\bibfield  {journal} {\bibinfo  {journal} {Phys. Rev. D}\ }\textbf {\bibinfo {volume} {37}},\ \bibinfo {pages} {2484} (\bibinfo {year} {1988})}\BibitemShut {NoStop}%
\bibitem [{\citenamefont {Caron-Huot}\ and\ \citenamefont {Moore}(2008{\natexlab{a}})}]{Caron-Huot:2007rwy}%
  \BibitemOpen
  \bibfield  {author} {\bibinfo {author} {\bibfnamefont {S.}~\bibnamefont {Caron-Huot}}\ and\ \bibinfo {author} {\bibfnamefont {G.~D.}\ \bibnamefont {Moore}},\ }\href {\doibase 10.1103/PhysRevLett.100.052301} {\bibfield  {journal} {\bibinfo  {journal} {Phys. Rev. Lett.}\ }\textbf {\bibinfo {volume} {100}},\ \bibinfo {pages} {052301} (\bibinfo {year} {2008}{\natexlab{a}})},\ \Eprint {http://arxiv.org/abs/0708.4232} {arXiv:0708.4232 [hep-ph]} \BibitemShut {NoStop}%
\bibitem [{\citenamefont {Caron-Huot}\ and\ \citenamefont {Moore}(2008{\natexlab{b}})}]{Caron-Huot:2008dyw}%
  \BibitemOpen
  \bibfield  {author} {\bibinfo {author} {\bibfnamefont {S.}~\bibnamefont {Caron-Huot}}\ and\ \bibinfo {author} {\bibfnamefont {G.~D.}\ \bibnamefont {Moore}},\ }\href {\doibase 10.1088/1126-6708/2008/02/081} {\bibfield  {journal} {\bibinfo  {journal} {JHEP}\ }\textbf {\bibinfo {volume} {02}},\ \bibinfo {pages} {081} (\bibinfo {year} {2008}{\natexlab{b}})},\ \Eprint {http://arxiv.org/abs/0801.2173} {arXiv:0801.2173 [hep-ph]} \BibitemShut {NoStop}%
\bibitem [{\citenamefont {Petreczky}\ and\ \citenamefont {Teaney}(2006)}]{Petreczky:2005nh}%
  \BibitemOpen
  \bibfield  {author} {\bibinfo {author} {\bibfnamefont {P.}~\bibnamefont {Petreczky}}\ and\ \bibinfo {author} {\bibfnamefont {D.}~\bibnamefont {Teaney}},\ }\href {\doibase 10.1103/PhysRevD.73.014508} {\bibfield  {journal} {\bibinfo  {journal} {Phys. Rev. D}\ }\textbf {\bibinfo {volume} {73}},\ \bibinfo {pages} {014508} (\bibinfo {year} {2006})},\ \Eprint {http://arxiv.org/abs/hep-ph/0507318} {arXiv:hep-ph/0507318} \BibitemShut {NoStop}%
\bibitem [{\citenamefont {Caron-Huot}\ \emph {et~al.}(2009)\citenamefont {Caron-Huot}, \citenamefont {Laine},\ and\ \citenamefont {Moore}}]{Caron-Huot:2009ncn}%
  \BibitemOpen
  \bibfield  {author} {\bibinfo {author} {\bibfnamefont {S.}~\bibnamefont {Caron-Huot}}, \bibinfo {author} {\bibfnamefont {M.}~\bibnamefont {Laine}}, \ and\ \bibinfo {author} {\bibfnamefont {G.~D.}\ \bibnamefont {Moore}},\ }\href {\doibase 10.1088/1126-6708/2009/04/053} {\bibfield  {journal} {\bibinfo  {journal} {JHEP}\ }\textbf {\bibinfo {volume} {04}},\ \bibinfo {pages} {053} (\bibinfo {year} {2009})},\ \Eprint {http://arxiv.org/abs/0901.1195} {arXiv:0901.1195 [hep-lat]} \BibitemShut {NoStop}%
\bibitem [{\citenamefont {Banerjee}\ \emph {et~al.}(2012)\citenamefont {Banerjee}, \citenamefont {Datta}, \citenamefont {Gavai},\ and\ \citenamefont {Majumdar}}]{Banerjee:2011ra}%
  \BibitemOpen
  \bibfield  {author} {\bibinfo {author} {\bibfnamefont {D.}~\bibnamefont {Banerjee}}, \bibinfo {author} {\bibfnamefont {S.}~\bibnamefont {Datta}}, \bibinfo {author} {\bibfnamefont {R.}~\bibnamefont {Gavai}}, \ and\ \bibinfo {author} {\bibfnamefont {P.}~\bibnamefont {Majumdar}},\ }\href {\doibase 10.1103/PhysRevD.85.014510} {\bibfield  {journal} {\bibinfo  {journal} {Phys. Rev. D}\ }\textbf {\bibinfo {volume} {85}},\ \bibinfo {pages} {014510} (\bibinfo {year} {2012})},\ \Eprint {http://arxiv.org/abs/1109.5738} {arXiv:1109.5738 [hep-lat]} \BibitemShut {NoStop}%
\bibitem [{\citenamefont {Brambilla}\ \emph {et~al.}(2020)\citenamefont {Brambilla}, \citenamefont {Leino}, \citenamefont {Petreczky},\ and\ \citenamefont {Vairo}}]{Brambilla:2020siz}%
  \BibitemOpen
  \bibfield  {author} {\bibinfo {author} {\bibfnamefont {N.}~\bibnamefont {Brambilla}}, \bibinfo {author} {\bibfnamefont {V.}~\bibnamefont {Leino}}, \bibinfo {author} {\bibfnamefont {P.}~\bibnamefont {Petreczky}}, \ and\ \bibinfo {author} {\bibfnamefont {A.}~\bibnamefont {Vairo}},\ }\href {\doibase 10.1103/PhysRevD.102.074503} {\bibfield  {journal} {\bibinfo  {journal} {Phys. Rev. D}\ }\textbf {\bibinfo {volume} {102}},\ \bibinfo {pages} {074503} (\bibinfo {year} {2020})},\ \Eprint {http://arxiv.org/abs/2007.10078} {arXiv:2007.10078 [hep-lat]} \BibitemShut {NoStop}%
\bibitem [{\citenamefont {Altenkort}\ \emph {et~al.}(2021)\citenamefont {Altenkort}, \citenamefont {Eller}, \citenamefont {Kaczmarek}, \citenamefont {Mazur}, \citenamefont {Moore},\ and\ \citenamefont {Shu}}]{Altenkort:2020fgs}%
  \BibitemOpen
  \bibfield  {author} {\bibinfo {author} {\bibfnamefont {L.}~\bibnamefont {Altenkort}}, \bibinfo {author} {\bibfnamefont {A.~M.}\ \bibnamefont {Eller}}, \bibinfo {author} {\bibfnamefont {O.}~\bibnamefont {Kaczmarek}}, \bibinfo {author} {\bibfnamefont {L.}~\bibnamefont {Mazur}}, \bibinfo {author} {\bibfnamefont {G.~D.}\ \bibnamefont {Moore}}, \ and\ \bibinfo {author} {\bibfnamefont {H.-T.}\ \bibnamefont {Shu}},\ }\href {\doibase 10.1103/PhysRevD.103.014511} {\bibfield  {journal} {\bibinfo  {journal} {Phys. Rev. D}\ }\textbf {\bibinfo {volume} {103}},\ \bibinfo {pages} {014511} (\bibinfo {year} {2021})},\ \Eprint {http://arxiv.org/abs/2009.13553} {arXiv:2009.13553 [hep-lat]} \BibitemShut {NoStop}%
\bibitem [{\citenamefont {Brambilla}\ \emph {et~al.}(2023)\citenamefont {Brambilla}, \citenamefont {Leino}, \citenamefont {Mayer-Steudte},\ and\ \citenamefont {Petreczky}}]{Brambilla:2022xbd}%
  \BibitemOpen
  \bibfield  {author} {\bibinfo {author} {\bibfnamefont {N.}~\bibnamefont {Brambilla}}, \bibinfo {author} {\bibfnamefont {V.}~\bibnamefont {Leino}}, \bibinfo {author} {\bibfnamefont {J.}~\bibnamefont {Mayer-Steudte}}, \ and\ \bibinfo {author} {\bibfnamefont {P.}~\bibnamefont {Petreczky}} (\bibinfo {collaboration} {TUMQCD}),\ }\href {\doibase 10.1103/PhysRevD.107.054508} {\bibfield  {journal} {\bibinfo  {journal} {Phys. Rev. D}\ }\textbf {\bibinfo {volume} {107}},\ \bibinfo {pages} {054508} (\bibinfo {year} {2023})},\ \Eprint {http://arxiv.org/abs/2206.02861} {arXiv:2206.02861 [hep-lat]} \BibitemShut {NoStop}%
\bibitem [{\citenamefont {Altenkort}\ \emph {et~al.}(2023)\citenamefont {Altenkort}, \citenamefont {Kaczmarek}, \citenamefont {Larsen}, \citenamefont {Mukherjee}, \citenamefont {Petreczky}, \citenamefont {Shu},\ and\ \citenamefont {Stendebach}}]{Altenkort:2023oms}%
  \BibitemOpen
  \bibfield  {author} {\bibinfo {author} {\bibfnamefont {L.}~\bibnamefont {Altenkort}}, \bibinfo {author} {\bibfnamefont {O.}~\bibnamefont {Kaczmarek}}, \bibinfo {author} {\bibfnamefont {R.}~\bibnamefont {Larsen}}, \bibinfo {author} {\bibfnamefont {S.}~\bibnamefont {Mukherjee}}, \bibinfo {author} {\bibfnamefont {P.}~\bibnamefont {Petreczky}}, \bibinfo {author} {\bibfnamefont {H.-T.}\ \bibnamefont {Shu}}, \ and\ \bibinfo {author} {\bibfnamefont {S.}~\bibnamefont {Stendebach}} (\bibinfo {collaboration} {HotQCD}),\ }\href {\doibase 10.1103/PhysRevLett.130.231902} {\bibfield  {journal} {\bibinfo  {journal} {Phys. Rev. Lett.}\ }\textbf {\bibinfo {volume} {130}},\ \bibinfo {pages} {231902} (\bibinfo {year} {2023})},\ \Eprint {http://arxiv.org/abs/2302.08501} {arXiv:2302.08501 [hep-lat]} \BibitemShut {NoStop}%
\bibitem [{\citenamefont {Altenkort}\ \emph {et~al.}(2024)\citenamefont {Altenkort}, \citenamefont {de~la Cruz}, \citenamefont {Kaczmarek}, \citenamefont {Larsen}, \citenamefont {Moore}, \citenamefont {Mukherjee}, \citenamefont {Petreczky}, \citenamefont {Shu},\ and\ \citenamefont {Stendebach}}]{Altenkort:2023eav}%
  \BibitemOpen
  \bibfield  {author} {\bibinfo {author} {\bibfnamefont {L.}~\bibnamefont {Altenkort}}, \bibinfo {author} {\bibfnamefont {D.}~\bibnamefont {de~la Cruz}}, \bibinfo {author} {\bibfnamefont {O.}~\bibnamefont {Kaczmarek}}, \bibinfo {author} {\bibfnamefont {R.}~\bibnamefont {Larsen}}, \bibinfo {author} {\bibfnamefont {G.~D.}\ \bibnamefont {Moore}}, \bibinfo {author} {\bibfnamefont {S.}~\bibnamefont {Mukherjee}}, \bibinfo {author} {\bibfnamefont {P.}~\bibnamefont {Petreczky}}, \bibinfo {author} {\bibfnamefont {H.-T.}\ \bibnamefont {Shu}}, \ and\ \bibinfo {author} {\bibfnamefont {S.}~\bibnamefont {Stendebach}} (\bibinfo {collaboration} {HotQCD}),\ }\href {\doibase 10.1103/PhysRevLett.132.051902} {\bibfield  {journal} {\bibinfo  {journal} {Phys. Rev. Lett.}\ }\textbf {\bibinfo {volume} {132}},\ \bibinfo {pages} {051902} (\bibinfo {year} {2024})},\ \Eprint {http://arxiv.org/abs/2311.01525} {arXiv:2311.01525 [hep-lat]} \BibitemShut {NoStop}%
\bibitem [{\citenamefont {Adamov\'a}\ \emph {et~al.}(2019)\citenamefont {Adamov\'a} \emph {et~al.}}]{Adamova:2019vkf}%
  \BibitemOpen
  \bibfield  {author} {\bibinfo {author} {\bibfnamefont {D.}~\bibnamefont {Adamov\'a}} \emph {et~al.},\ }\href@noop {} {\  (\bibinfo {year} {2019})},\ \Eprint {http://arxiv.org/abs/1902.01211} {arXiv:1902.01211 [physics.ins-det]} \BibitemShut {NoStop}%
\end{thebibliography}%

\end{document}